\documentclass[aps,prd,showpacs,twocolumn,nofootinbib,superscriptaddress,amsmath,amssymb]{revtex4}
\usepackage{amsfonts}
\usepackage{bm}
\usepackage{verbatim}

\usepackage{graphicx}
\usepackage{graphics}

\usepackage{color}

\newcommand{\revision}[1]{{#1}}

\usepackage{epsfig}

\begin{document}

\title{Hall effect in the presence of rotation}

\author{M.A. Zubkov \footnote{On leave of absence from Institute for Theoretical and Experimental Physics, B. Cheremushkinskaya 25, Moscow, 117259, Russia}}
\email{zubkov@itep.ru}
\affiliation{Physics Department, Ariel University, Ariel 40700, Israel}

\begin{abstract}
The rotating relativistic fermion system is considered. The consideration is based on the Dirac equation written in the laboratory (non - rotating) reference frame. Rotation in this approach gives rise to the effective magnetic and electric fields that act in the same way both on positive and negative electric charges. In the presence of external electric field in the given system the electric current appears orthogonal to both the electric field and the axis of rotation. The possible applications to the physics of quark - gluon plasma are discussed.
\end{abstract}

\maketitle

\section{Introduction}
\label{SectIntro}

Rotation of the system of relativistic fermions is encountered in many situations. The particular examples are the rotated neutron stars~\cite{Cook:1993qr}, and the rotated fireballs that appear in the non - central heavy ion collisions~\cite{ref:HIC}. In those cases the quark matter may exist in the exotic phases of QCD such as the quark - gluon plasma phase \cite{QCDphases,1,2,3,4,5,6,7,8,9,10}. Sure, the rotated plasma consisted of the colorless particles (hadrons and leptons) also appears in a variety of cases. In particular, the rotating neutrino systems were discussed in  ~\cite{Vilenkin:1980zv,ref:Vilenkin}. Various anomalous transport phenomena are specific for rotation~\cite{ref:CVE}. It is also worth mentioning that there is a certain relation of this class of phenomena with the solid state physics   ~\cite{ref:Weyl}.

Rotation of free fermions has been investigated in a number of papers (see, for example, ~\cite{Iyer:1982ah,ref:Becattini,Ambrus:2014uqa,Ambrus:2015lfr,Manning:2015sky} and references therein). For the case of the interacting systems rotation affects the chiral symmetry breaking \cite{ref:McInnes,Chen:2015hfc,Jiang:2016wvv,Ebihara:2016fwa,Chernodub:2016kxh}. The rigidly rotating system may be considered only in a region of space of limited size because sufficiently far from the rotation axis velocity of the substance exceeds the speed of light~\cite{Ambrus:2014uqa,ref:superluminal,Chernodub:2017ref}. It is also interesting to consider the rotating system in the presence of magnetic field directed along the rotation axis~\cite{Chen:2015hfc}.

Unlike the case of a system staying at rest as a whole, for the case of rotation one might encounter an ambiguity of the definition of vacuum. In order to understand such an ambiguity let us consider the solid, the electronic system of which simulates relativistic physics, that is the corresponding excitations are described by the Dirac equation at low energies. This is the case of the so - called Dirac insulator or the cases of Dirac or Weyl semimetals \cite{semimetal_discovery,semimetal_discovery2,semimetal_discovery3,semimetal_effects6,semimetal_effects10,semimetal_effects11,semimetal_effects12,semimetal_effects13,Zyuzin:2012tv,tewary,16}. (In the cases of Dirac and Weyl semimetals the mass term in Dirac equation vanishes.) Dirac vacuum in the solid is the set of occupied energy levels placed below zero. The quasiparticle excitations correspond to the occupied states above zero (particles) and to the vacant states below zero (antiparticles). The similar situation takes place for the quantum liquids \cite{Volovik2003}. In particular, in the superfluid $^3$He-A the superfluid component of liquid represents vacuum (i.e. the occupied states below zero) while the excitations (described by Dirac equation) correspond to the normal component.
In \cite{VZ} (Remark 2.1.) the possible existence of the new type of Weyl points in multi - fermion systems of general type was noticed. Later the corresponding materials were called the type II Weyl semimetals and discovered experimentally \cite{W2}. In these materials the Dirac cone is overtilted, so that instead of the Fermi point the two Fermi pockets appear surrounded by the Fermi surface. It has been proposed, that in those novel materials the fermionic quasiparticles behave similar to the fermions inside the horizon of the equilibrium black holes \cite{VolovikBH,VolovikBHW2}.

Rotation in the mentioned above systems may be considered in the two ways: either the solid body (or a liquid) is rotated as a whole, or the solid body (the superfluid component of liquid) stays at rest and only the quasiparticles are rotated. The first case is typical for the experiments with real solids while the second case is typical for the laboratory experiments with quantum liquids. In the latter case typically a container is rotated, which gives rotation to the normal component ($=$ quasiparticles), while the superfluid component may remain at rest.

One can easily see, how different are the two physical situations described above for the solids and liquids. Since those systems may be considered as analogue computers for the simulation of the physics of elementary particles, one might expect certain differences in the behavior of the quantum field theory when vacuum is staying at rest and when vacuum itself is rotated. If the difference is indeed present, then the second version of the field theory looks not appropriate for the description of the rotated fireballs of the quark - gluon plasma created during the heavy - ion collisions.

The non-rotating vacuum has been considered previously, in particular, in \cite{Vilenkin:1980zv}. In the majority of mentioned above papers, however, rotation was considered in the rotating reference frame (see for example \cite{Iyer:1982ah,Ambrus:2014uqa,Ambrus:2015lfr}. For the reasons discussed above we do not use the approach with the description of the system in the rotated reference frame.  Instead, we work out the model, in which vacuum stays at rest, and the rotation of quasiparticles is ensured by the presence of the term in the action, which generalizes the usual chemical potential term. This approach to the system of particles with the macroscopic motion has been proposed for the first time in \cite{chiralhydro}.

We will discuss in the framework of this approach the particular effect that takes place in the rotating relativistic plasma. This is the appearance of the electric current in the presence of the weak external electric field orthogonal to the rotation axis. We will demonstrate, that this current is orthogonal both to the electric field and to the rotation axis. Therefore, we may speak of the kind of the Hall effect.

The classical Hall effect appears in the system of classical charged particles placed in the magnetic and electric fields orthogonal to each other \cite{Hall}. The origin of this effect is the force acting on the particle of charge $Q$ moving with the velocity ${\bf v}$ from the side of the electromagnetic field with the strength $({\bf E},{\bf H})$:
$$
{\bf F} = Q\, {\bf E} + Q \,[{\bf v}\times {\bf H}]
$$
If ${\bf E} \bot {\bf H}$ and $|{\bf E}| < |{\bf H}|$ we are able to consider the particle in the reference frame moving with velocity ${\bf V}=\frac{{\bf E}\times {\bf H}}{{\bf H}^2}$. In this reference frame the electric field vanishes, and the equation of motion is $\frac{d}{dt}{\bf p}^\prime = Q [{\bf v}^\prime \times {\bf H}^\prime]$, where ${\bf v}^\prime$ and ${\bf H}^\prime$ are the quantities in the new reference frame. The corresponding motion is localized within a finite region of space. Therefore, in the initial reference frame the particles drift with velocity ${\bf V}$ and produce electric current directed orthogonally to the electric field.

In strong magnetic field at low temperatures, when quantum mechanics becomes important, the quantum Hall effect appears in the two dimensional systems \cite{QH}. Its essence is similar to that of the classical Hall effect, but acquires the new features. In the clean systems instead of the classical trajectory in the moving reference frame the Landau levels appear. The overall motion in the direction orthogonal to the electric field leads to the Hall current that is quantized because only those states are involved into the formatio of current, that have energies below the Fermi level, while the energies are quantized. In the presence of disorder the effect remains but the Hall current is distributed in a nontrivial way between the bulk and the boundary. Under certain circumstances the whole current flows close to the boundaries.

In certain $2D$ quantum systems the so - called anomalous quantum Hall effect takes place \cite{AQHE}. Its essence is the appearance of electric current orthogonal to the electric field without any external magnetic field. The role of magnetic field is played here by the nontrivial topology of momentum space \cite{Volovik2003}. Recently the presence of this effect was also predicted in the three -- dimensional systems: the Weyl semimetals \cite{Zyuzin:2012tv,tewary} and topological insulators \cite{Wigner}.

In the present paper we announce the new version of quantum Hall effect, which occurs in the true relativistic $3+1$ D systems. The role of the magnetic field is played here by rotation. Namely, we will argue that in the plasma of electrically charged particles in the presence of rotation and the weak electric field (directed orthogonally to the rotation axis) the electric current appears directed orthogonally both to the rotation axis and to the electric field. We will also point out the possibility to observe this effect in the rotating fireballs of quark - gluon plasma that appear in the heavy - ion collisions \cite{ref:HIC}. It is worth mentioning, that the Hall effect has been considered recently in the dense baryonic matter in the presence of rotation using the different approach in \cite{Huang:2017pqe}. \revision{The similar phenomena have been considered also in \cite{Baumgartner:2018dqi}.}

\section{Free massive fermions in rotating medium}
\label{SecFree}

In this section we consider the fermionic matter rotating with angular velocity \(\Omega\) depending on $r$  in nearly equilibrium state with the chemical potential \(\mu\) that also depends on $r$. First we neglect the interaction between the fermions. It will, however, not alter the results presented here as will be explained below. We require that the dependence of $\Omega$ on $r$ provides that the velocity $\Omega r$ never becomes larger than $1$. Say, we may consider the following form of dependence:  $
  \Omega(r) = \Omega_0 \frac{1}{1+r/R}
  $,
 where $\Omega$ and $R$ are the constants that obey $\Omega_0 R <1$.
Performing the Lorentz boost to the rest frame in the small vicinity of a given point one gets \cite{chiralhydro} the action
\begin{equation}
        S = \int d^4x \bar{\psi}(\gamma^\mu(i\partial_\mu+\mu u_\mu)-M)\psi\label{action}
\end{equation}
where \(u_\mu\) is the four-velocity of the medium in the given point, which is uniquely defined as a unit four-vector tangent to a world-line of a piece of medium.  It can be obtained from the ordinary velocity of the piece of medium \(\vec{v}=(-\Omega(r) y,\Omega(r) x,0)^T\) as follows
    \begin{equation}
        u^\mu=\frac{dx^\mu}{ds}=\frac{(dt,\vec{v}dt)}{ds}
    \end{equation}
\(ds\) is the Lorentz-invariant interval and it is equal to time increment in an inertial frame, in which the piece of medium rests at the moment.
    So \(ds=dt\sqrt{1-v^2}\) and
    \begin{equation}
        u^\mu = \gamma(r)(1,-\Omega(r) y,\Omega(r) x,0)^T,\quad \gamma(r)=\frac{1}{\sqrt{1-\Omega^2(r)r^2}}
    \end{equation}
The corresponding Dirac equation is precisely equivalent to the Dirac equation in the presence of the external Abelian gauge field potential $A_\mu = - \mu u_\mu$:
    \begin{gather}
        (\gamma^\mu(i\partial_\mu+\mu u_\mu)-M)\psi=0 \label{direq}
     \end{gather}
The space components of $u_\mu$ are equivalent to the effective magnetic field, which is directed along the axis of rotation and depends on $r$:
\begin{eqnarray}
    \mathbf{B} &=& -(\nabla\times(\mu(r)\mathbf{u}(r))) \nonumber\\&& = -\mu(r) \Omega(r) \gamma(r) \Big(2+  \frac{d\,{\rm log}\,[ \mu(r) \Omega(r) \gamma(r)]}{d\,{\rm log}\,r}\Big)\mathbf{e}_z \label{Magn}
\end{eqnarray}
The $0$ - component of $u_\mu$ gives rise to the electric field with radial direction
\begin{gather}
    \mathbf{E} = \nabla \mu(r) u_0(r) = \frac{d (\mu(r)\gamma(r))}{dr}\frac{\mathbf{r}}{r}
\end{gather}
The particularly simple case is when the chemical potential in the laboratory reference frame  is constant $$\mu_{lab} = \mu(r) \gamma(r) = const$$ Then we obtain $\mathbf{E} = 0$ while \begin{equation}
\mathbf{B} = (0,0, -2\mu_{lab}\Omega(r)  [1+\frac{1}{2} \frac{d {\rm log}\,\Omega(r) }{d\,{\rm log}\,r}]) \label{Beff}\end{equation}
In the following we will restrict ourselves by this case. \revision{This approach to the description of macroscopic motion is relevant if velocity is slowly varying. To find the corresponding conditions we compare the effective magnetic field with the mass squared:  $|\mathbf{B}| \sim 2 \mu_{lab} \Omega \ll M^2$.} If we would require, in addition, that in the considered region of space $\Omega$ is constant, then we are left with the Dirac fermion in the presence of constant effective magnetic field $-2\mu_{lab} \Omega$. \revision{The corresponding spectrum is discrete (at each value of momentum component $p_z$ along the axis of rotation) and consists of the Landau levels (see, for example, \cite{9}):
\begin{equation}
{\cal E} = \pm \sqrt{p_z^2 + M^2 + 2 n |\mathbf{B}|}, \quad n = 0, 1, ...\label{LL}
\end{equation}}
\revision{It is worth mentioning, that there are several different definitions of the macroscopic motion of the substance described by the quantum field theory. Here we chose the definition of rotation based on the local boost following \cite{chiralhydro}. Previously most of the papers utilized the approach based on the enhancement of angular momentum (see, for example, \cite{Vilenkin:1980zv}) and on the global transformation to the rotated reference frame (see, for example, \cite{Iyer:1982ah,Ambrus:2014uqa,Ambrus:2015lfr,Manning:2015sky,Chen:2015hfc}). In \cite{Chernodub:2016kxh} it has been demonstrated that those two approaches give identical results for the appropriate choice of the boundary conditions. Also it has been shown, that the energy levels in the presence of rotation are similar to the Landau levels (see, for example, Eq.(2.24) of \cite{Chernodub:2016kxh}). Therefore, this is not a surprise, that using the approach of \cite{chiralhydro} to the description of rotation we also obtain the Landau levels in the case of the constant chemical potential. However, quantitatively Eq. (\ref{LL}) differs from Eq. (2.24) of \cite{Chernodub:2016kxh}. This difference reflects the fact that the macroscopic motion of substance in the quantum field theory may be described in a variety of different ways.  }

\section{Macroscopic rotation in the presence of external electric field}

In this section we consider the rotated system described by the effective action Eq. (\ref{action}) in the presence of external electric field ${\bf E}_{ext}$ normal to the rotation axis. We discuss here the case, when \revision{$M^2 \gg |2\mu\Omega| \gg |E_{ext}|$} while $\mu_{lab} = \mu(r) \gamma(r) = const$. The effective gauge potential has the form
\begin{equation}
A^\mu = (-\mu_{lab}-E_{ext}x,\mu_{lab} \Omega(r) y,-\mu_{lab}\Omega(r) x,0)^T \label{Aeff}
\end{equation}
that gives rise to the magnetic field of Eq. (\ref{Beff}) and electric field
\begin{gather}
\mathbf{E} = (E_{ext},0,0)
\end{gather}
Our further assumption is that the rotation velocity varies slowly, i.e. its change is small on the scale of the magnetic length $l_B = 1/\sqrt{B}$. \revision{Together with the condition for the applicability of Eq. (\ref{action}) we write it as
\begin{equation}
l_B M \gg 1,\quad \frac{d l_B}{d r}\sim \frac{d (1/\sqrt{\mu_{lab}\Omega})}{dr} \ll 1 \label{magnlength}
\end{equation}}

In this case the states in a small vicinity of the particular point $r_0$ correspond to the Landau levels in the presence of the constant magnetic field. Those states drift locally along the equipotential lines of the electric potential, that are the straight lines parallel to the $y$ axis. (There is also the motion of fermions in the direction of the rotation axis $z$. But it does not give the overall motion of the substance as a whole.) The velocity of the drift is given by
\begin{equation}
{\bf V} = \frac{{\bf E}_{ext}\times {\bf \Omega}}{2 \mu_{lab} \Omega^2}\,
\end{equation}
In the reference frame moving with this velocity the effective electric field vanishes.
In the original reference frame the substance as a whole moves with this velocity in the direction orthogonal to both the rotation axis and the external electric field.

In the above consideration we neglected completely the interactions between the fermions due to the exchange by the gauge bosons. In order to take them into account we should add to the action of Eq. (\ref{action}) the interaction term. The result of the integration over the gauge field is the effective action, which may be rather complicated, but still it describes the fermion system in the presence of the effective external gauge field of Eq. (\ref{Aeff}). As for the non - interacting system we come locally to the reference frame moving with velocity $\bf V$, where the effective electric field  vanishes.
In the new reference frame the effective fermion action may be rather complicated. It does not contain the electric field but contains the effective magnetic field. The spectrum of quasiparticles may differ essentially from the ordinary Landau levels  due to the interactions. However, according to the Bloch theorem the presence of the magnetic field only cannot cause the macroscopic motion (see \cite{Yamamoto} and references therein). Although this theorem has been proved completely for the non  - relativistic case only, we expect that it remains valid in the relativistic case as well. The known particular relativistic  cases (see, for example, \cite{Z2016}) confirm this supposition.  Therefore, we suppose that in the reference frame moving with velocity $\bf V$ the substance remains at rest, and therefore it moves locally with velocity $\bf V$ in the laboratory reference frame.

\section{Electric current in the rotating plasma}

Let us consider the case of plasma that consists of the fermions of several types with positive and negative charges. For each fermion we have its own chemical potential $\mu_f$ that gives rise to the nonzero density $\rho_f$. Let us start from the non - interacting case. The interactions do not alter essentially our conclusions. We suppose, that inside the substance there is the region, where the rotation takes place with the angular velocity that decreases far from the rotation axis. This setup (with the interactions added) may be acceptable for the qualitative consideration of the rotated fireball that may appear in the non - central heavy - ion collisions.

The fireball cannot be stable and it decays into the jets of hadrons. However, during a certain period of time the fireball may be considered as existing in a quasi - equilibrium state characterized by temperature and chemical potentials $\mu_f$. Following the previous sections we restrict ourselves by the case when the chemical potential in the laboratory reference frame is constant. We omit the subscript $lab$ for brevity and denote by $\mu_f$ the chemical potential of the given constituent of the plasma in the laboratory reference frame. As for the angular velocity, we assume that it obeys the condition of Eq. (\ref{magnlength}) that the effective magnetic length varies slowly.

Let us add the external electric field orthogonal to the rotation axis.
We have shown above, that the system of charges $Q_f$ moves as a whole with local velocity
\begin{equation}
{\bf V}_f  =  \frac{{\bf E}_{ext}\times {\bf \Omega}}{2 \mu_{f} \Omega^2}Q_f\,\label{vf}
\end{equation}
Let us suppose for the moment that the charges $Q_f$ interact with each other but cannot interact with the particles that have different values of electric charges. Then in the reference frame moving with velocity ${\bf V}_f$ of Eq. (\ref{vf}) the total $4$ - current is given by
$$
j^\prime_f = Q_f(\rho^\prime_f,0)
$$
where $\rho_f^\prime$ is the density of particles in the moving reference frame. In the original reference frame the electric current is given by Lorentz transformation:
$$
{\bf j}_f = Q_f \frac{\rho^\prime_f {\bf V}_f}{\sqrt{1-{\bf V}_f^2}} = Q_f \rho_f {\bf V}_f =   \frac{\rho_f}{\mu_{f}} Q_f^2\frac{{\bf E}_{ext}\times {\bf \Omega}}{2 \Omega^2} $$
If there would be no interactions between the particles of different values of electric charge, then we obtain:
\begin{equation}
{\bf j} = \sum_f \frac{\rho_f}{\mu_{f}} Q_f^2\frac{{\bf E}_{ext}\times {\bf \Omega}}{2 \Omega^2}\label{jt}
\end{equation}
The presence of the interactions between the particles with different values of $Q_f$ qualitatively may be understood using an analogy with the quantum Hall effect in solids. Namely, the collection of the particles with charges $Q_{f_2}$ plays for the system of particles with charges $Q_{f_1}$ the role similar to that of the impurities for the electron quasiparticles. The remarkable feature of the quantum Hall effect is that the impurities do not affect the total Hall conductance. Though, the density of the resulting current is rearranged in such a way, that it may be concentrated close to the boundary. Therefore, we propose the hypothesis that the effective values of velocities ${\bf V}_f$ are not changed when the interactions between different kinds of fermions are included even if those interactions are strong.  Thus, we expect that Eq. (\ref{jt}) gives the total electric current for the case when  all interactions are included. However, do not exclude that the resulting electric current calculated using this expression may actually be concentrated close to the boundary of the system.

Here and below we assume that $\Omega \ll \mu_f, T$ so that we are able to neglect the effect of rotation on the dependence of $\rho_f$ on $\mu_f$. Such values of $\Omega$ are typical for the rotating fireballs that appear in the non - central heavy ion collisions (see, for example, \cite{ref:HIC}). Besides we imply that $\mu_f$ and $T$ may be considered as constant within the given region of space while $\mu_f, T  \gg M$.
In QED we have the following expression for the $\mu_f$ - dependent terms in the thermodynamical potential for each type of quarks
 \cite{Toimela:1984xy}:
$$
\Omega^{QED}_f = - \Big(\frac{\mu_f^4}{12 \pi^2}+\frac{\mu_f^2T^2}{6}\Big) + \frac{\alpha}{4\pi} \Big(\mu_f^2T^2+\frac{\mu_f^4}{2\pi^2} \Big)
$$
This gives the electric current
\begin{equation}
{\bf j}^{QED} =   \sum_f \frac{1}{3\pi^2}\Big(\mu^2_f + \pi^2 T^2\Big)\Big[1-\frac{3}{2}\frac{\alpha}{\pi}\Big]Q^2_f\frac{{\bf E}_{ext}\times {\bf \Omega}}{2 \Omega^2}
\end{equation}
For the pure QED we obtain the following Hall conductivity:
\begin{equation}
\sigma^{QED}_H =    \sum_f\frac{1}{6\pi^2}\frac{\Big(\mu^2_f + \pi^2 T^2\Big)}{\Omega_0}\Big[1-\frac{3\alpha}{2 \pi}\Big]Q^2_f
\end{equation}

The case of the quark - gluon plasma, when the particles are charged with respect to the color gauge field is much more involved. The strong interactions dominate, and at the asymptotically large temperatures the Hard Thermal Loop (HTL) approximation may be applied. It gives already in the leading order \cite{Andersen:2012wr} a rather complicated expression, which is given below for completeness:
\begin{eqnarray}
\Omega^{(HTL)QCD} &=& \frac{d_A \pi^2 T^4}{45}\,\Big(\frac{30 N_c}{d_A}\sum_f(\hat{\mu}_f^2+2\hat{\mu}_f^4)-\frac{15}{2}\hat{m}_D^2\nonumber\\&&
-\frac{30N_c}{d_A}\sum_f (1+12 \hat{\mu}_f^2) \hat{m}_{q,f}^2 + 30 \hat{m}_D^3\nonumber\\&& + \frac{60 N_c}{d_A}(6-\pi^2)\sum_f\hat{m}^4_{q,f}\nonumber\\&& + \frac{45}{4}(\gamma_E - \frac{7}{2} + \frac{\pi^2}{3}+{\rm log}\,\frac{\bar{\Lambda}}{4 \pi T})\hat{m}_D^4\Big)
\end{eqnarray}
Here $\gamma_E$ is the Euler constant while the thermal fermion masses $m_{q,f}$ and the Debye mass $m_D$ are given by
\begin{eqnarray}
m^2_D & = & \frac{g^2}{3}\Big[(N_c + \frac{N_f}{2})T^2+\frac{3}{2\pi^2}\sum_f\mu_f^2 \Big]\nonumber\\
m^2_{q,f} & = & \frac{g^2}{4}\frac{N_c^2-1}{4N_c}(T^2+\frac{\mu_f^2}{\pi^2})
\end{eqnarray}
where $N_c = 3$ is the number of colors.
Another notations used above are:
\begin{eqnarray}
\hat{m}_{q,f} &=& \frac{m_{q,f}}{2\pi T}, \quad \hat{m}_{D} = \frac{m_{D}}{2\pi T}\nonumber\\ d_A &=& N_c^2 -1
\end{eqnarray}
By $N_f$ we denote the number of fermions that participate in the dynamics at the given values of temperature, that is we should take the number of quarks with masses smaller than $T$.  By $\bar{\Lambda}$ we denote the renormalization scale, which is to be taken $\bar{\Lambda} \approx 1.445 \times 2\pi T$ according to \cite{Andersen:2012wr}.
We expect that at the asymptotically large values of temperature the above expression for the HTL thermodynamical potential represents a reasonable approximation to the real thermodynamical potential of the quark  - gluon plasma, and we are able to estimate the effective electric current:
\begin{equation}
{\bf j}^{(HTL)QCD} =   \sum_f \frac{\partial \Omega^{(HTL)QCD}}{\partial \mu^2_f}Q^2_f\frac{{\bf E}_{ext}\times {\bf \Omega}}{ \Omega^2}
\end{equation}
We obtain the following estimate for the value of the Hall conductivity of the rotated quark - gluon plasma:
\begin{equation}
\sigma^{(HTL)QCD}_H =  \frac{1}{\Omega}  \sum_f\frac{\partial \Omega^{(HTL)QCD}}{\partial \mu^2_f}Q^2_f\label{condQCD}
\end{equation}
Recall, that Eq. (\ref{condQCD}) has been derived under the supposition that the angular velocity does not affect relation of the fermion densities and the chemical potentials. That means, that $\Omega \ll \mu_f, T$.

At the values of temperature just above the confinement - deconfinement transition we should take into account the nonperturbative contributions, which may change essentially the above result for the Hall conductivity. This, however, is out of the scope of the present paper.

\section{Conclusions}

In the present paper we consider the rotating system of relativistic fermions of several types characterized by the chemical potentials $\mu_f$, temperature, and the angular velocity. In this way the variety of physical problems with relativistic fermions may be described. In particular, qualitatively the obtained results may be applied to the description of the rotating fireballs that  appear in the non - central heavy - ion collisions. Such a fireball is unstable and decays into the jets of hadrons. However, during a certain short period of time the fireball may be considered approximately as the quasi - equilibrium state characterized by $\mu_f$, $\Omega$, and $T$. The quarks in the fireball possess several types of motion. First of all this is the rotation itself with the given angular velocity $\Omega$. Next, this is the motion that is caused by  interactions between the constituents of the fireball.

The key observation of the present paper is that if such a system is placed in the external electric field $E_{ext}$ orthogonal to the axis of rotation, and obeying $E_{ext} \ll \mu \Omega$, then there is an extra type of collective motion. The system of electrically charged particles moves as a whole in the direction orthogonal both to the rotation axis and to the direction of the electric field. The positive charges move in the direction opposite to that of the negative charges. As a result the electric current appears. Therefore, we are able to refer to it as to the Hall current caused by rotation. The appearance of this current may be observed in the behavior of the outgoing hadrons. Roughly speaking, the observer should find different number of positively and negatively charged hadrons moving in the opposite directions orthogonal to the rotation axis and to the direction of applied external electric field.

Unfortunately, at the present moment the quantitative investigation of the parameters ($\mu_f,\Omega,T$) describing the rotating fireballs is still in its infancy. A lot of efforts is still to be invested in order to give such a description. This is certainly out of the scope of the present paper. Only after such a description is given, it will be possible to calculate the Hall conductivity using Eq. (\ref{condQCD}). In the meantime we are able to conclude, that being placed in the extra electric field (orthogonal to the rotation axis) the fireball will give rise to the jets with an asymmetry of electric charge. Dealing with the collisions of heavy ions we may add such an electric field in the direction of the beam. Then it will be orthogonal to the rotation axis of the fireball. The orientation of the fireball may be reconstructed from the data on the distribution of the produced hadrons \cite{ref:HIC}. The distribution of electric charge of the hadrons should carry an information about the rotational Hall effect.

\revision{The conventional Hall effect is non - dissipative because the corresponding motion is orthogonal to the electric field. As long as the description by the slow varying effective magnetic field of the rotated system is accepted, the rotational Hall effect also may be considered as non - dissipative. It is worth mentioning, however, that the dissipation may appear, when corrections to Eq. (\ref{magnlength}) are taken into account. Then motion of the substance may acquire contributions along the external electric field. Those contributions should be accompanied by dissipation.}

The rotational Hall effect proposed here complements the family of the non - dissipative effects that may be observed in the heavy ion collisions. This family includes, in particular, the chiral vortical effect, the chiral separation effect and, possibly, the non - equilibrium version of the chiral magnetic effect (see, for example, \cite{Kharzeev} and references therein). On the technical side we use for the description of the rotation of medium the approach proposed in \cite{chiralhydro}. Within this approach the product of chemical potential $\mu$ and the macroscopic 4 -- velocity $u_\mu$ enters the action of the fermions as an effective electromagnetic 4 -- potential. Recall, that using this approach the current of the chiral vortical effect is derived easily. It appears as the current of the chiral separation effect (see, for example, \cite{Khaidukov:2017exf} and references therein), into which we substitute the effective magnetic field strength $F_{\mu\nu} = -\mu \partial_{[\mu} u_{\nu]}$. We obtain for the axial current along the rotation axis $j^\mu_5 = -\frac{\mu u_0}{4\pi^2} \epsilon^{\mu\nu\rho 0}F_{\nu\rho} = \frac{\mu^2}{2\pi^2} \epsilon^{\mu\nu\rho \sigma}u_\sigma\partial_{\nu} u_{\rho}$. Here $\mu_{lab} = \mu u_0$ is the chemical potential in the laboratory reference frame while $\mu$ is in the co - moving one. This is just the conventional chiral vortical effect \cite{ref:CVE}.

The author kindly acknowledges useful discussions with M.N.Chernodub, Z.V.Khaidukov and R.A.Abramchuk.

\end{document}